\documentclass{article}
\usepackage[utf8]{inputenc}
\usepackage[english]{babel}
\usepackage{amsmath}
\usepackage{amsfonts}
\usepackage[hyphens]{url}
\usepackage[breaklinks]{hyperref}
\usepackage{enotez}
\usepackage{todonotes}
\usepackage{amssymb}
\usepackage{lmodern}
\usepackage{geometry}
\usepackage{physics}
\usepackage{dsfont}
\usepackage{tabularx}
\usepackage{tikz}
\usetikzlibrary{decorations.pathmorphing}
\usepackage{relsize}
\usepackage{wrapfig}
\usepackage{csquotes}
\usepackage[authordate,natbib,isbn=false]{biblatex-chicago}
\AtEveryBibitem{\clearfield{note}}
\AtEveryCitekey{\clearfield{note}}
\AtEveryBibitem{\clearfield{month}}
\AtEveryBibitem{\ifentrytype{unpublished}{}{\clearfield{url}}}
\AtEveryBibitem{\clearfield{urlyear}}
\AtEveryBibitem{\clearfield{urlmonth}}
\AtEveryBibitem{\clearfield{urlday}}
\AtEveryCitekey{\clearfield{urlyear}}
\usepackage{graphicx}
\title{Spacetime from Entanglement: The Emergence of Metric, Gravity, or Topology}
\author{Rasmus Jaksland \\ Princeton University, Department of Philosophy, Princeton, New Jersey, USA}
\date{}

\begin{document}
\maketitle

\begin{abstract}
    In AdS/CFT, one often finds claims along the lines that ``spacetime emerges from entanglement." This paper argues that behind these general statements hide three distinct emergence claims about, respectively, metric, gravitational dynamics, and topological connectivity. Thus, despite being advertised with the same terminology, these results are not about the same spatiotemporal aspects. They can therefore not just be grouped as evidence for one unanimous conclusion, though they do point in similar directions. The paper also investigates whether the three emergence claims satisfy two of the necessary conditions for emergence: Determination and novelty. The paper argues that none of the emergence claims satisfies the determination condition: More than entanglement is needed to furnish the emergence basis. Besides entanglement, the emergence basis for the bulk metric must include the induced metric on the boundary. Thus, this claim might not satisfy the novelty condition for emergence that the emergent should be novel as compared to the emergence basis. Likewise, the emergence basis for topological connectivity seems to include connectivity whereby this emergence claim is also questionable. The paper concludes that only gravitational dynamics is novel compared to the fully furnished emergence basis.
\end{abstract}

\section{Introduction}
Under the title ``Spacetime from Entanglement," \citet{swingle_spacetime_2018} reviews recent results from the AdS/CFT correspondence and finds them to indicate that, ``[r]oughly speaking, entanglement is the fabric of spacetime" \citep[346]{swingle_spacetime_2018}, referring to it later as ``the emergence of spacetime from entanglement" \citep[356]{swingle_spacetime_2018}. 

The AdS/CFT correspondence states that there is a duality between certain theories defined on ($d+1$)-dimensional asymptotically anti de-Sitter (AdS) spacetimes (known as the bulk) and certain conformal field theories defined on the $d$-dimensional, asymptotic boundary of the dual spacetime (known as the boundary). The central result that relates spacetime and entanglement is the Hubeny-Rangamani-Takayanagi (HRT) formula \citeyearpar{hubeny_covariant_2007} which states that entanglement on the CFT boundary, as quantified by entanglement entropy, is proportional to areas in the AdS bulk (see Section \ref{Sec:HRT} for further details). Entanglement on the CFT side is, in other words, related to a feature of the geometry on the AdS side, and this serves as an important starting point for Swingle.

Swingle, however, is far from alone in taking the HRT formula to indicate that spacetime emerges from entanglement. Speaking in terms of ``geometry" instead of spacetime, which is also often seen in the literature, \citet{bao_towards_2019}, whose results will be discussed in some detail later, likewise notice that ``[t]he Ryu–Takayanagi and Hubeny–Rangamani–Takayanagi formulae suggest that bulk geometry emerges from the entanglement structure of the boundary theory" (\citealp[1]{bao_towards_2019}; see Section \ref{Sec:HRT} for more examples).

The philosophical literature has already questioned whether the AdS/CFT correspondence displays the kind of metaphysical asymmetry necessary for emergence \citep{rickles_ads/cft_2013,teh_holography_2013,dieks_emergence_2015}. This worry also applies to the specific claim that spacetime on the AdS side emerges from entanglement on the CFT side \citep[88]{ney_quantum_2021}. This paper will, however, discuss some other aspects of this alleged emergence of spacetime from entanglement that remain unclear in the recent philosophical discussions of this relation between spacetime and entanglement in AdS/CFT \citep[see, e.g.,][]{bain_rt_2021,jaksland_entanglement_2021,ney_quantum_2021}: What spacetime means in this context; whether entanglement alone is sufficient to furnish an emergence basis for spacetime; and, if not, whether spacetime remains sufficiently novel compared to the fully furnished emergence basis. 

The paper finds that at least three distinct emergence claims are advanced by the statements in the physics literature that spacetime emerges from entanglement. These are claims about the emergence of the bulk metric (Section \ref{Metric}), bulk gravitational dynamics (Section \ref{Gravitational}), and bulk topological connectivity (Section \ref{Topology}). The point here is not to argue that these ambiguities are examples of philosophical carelessness in the physics literature. Indeed, the context typically rather clearly reveals which specific spatiotemporal aspect that the emergence claim concerns. That claims about the emergence of spacetime from entanglement should be understood in their context like this is rather a note of caution to those philosophers and physicists alike who are interested in the big picture consequences of these results. This echoes the general warning of \citet{jaksland_many_2023} that `spacetime' is a rather ambiguous term and that investigations of the consequences of spacetime emergence are therefore better conducted in terms of specific spatiotemporal aspects. Indeed, the present paper shows that, if one begins such an investigation from the widespread claim that spacetime emerges from entanglement, then one will fail to appreciate that behind this apparent unanimity hides different emergence claims of rather differing degree of plausibility.

Investigating the different emergence claims in their specificity is what facilitates this assessment of whether these claims satisfy the conditions for emergence. More particularly, the focus will be on a determination condition and a novelty condition that Section \ref{Emergence} argues are necessary for emergence (as this term in used in physics). With respect to determination, the paper finds that, for the emergence of metric, gravitational dynamics, and topological connectivity alike, more than boundary entanglement is needed in the emergence basis. These emergence claims are a kind of contextual emergence claim, where some of the emergence basis is left implicit, what will be denoted `contrasting' emergence claims. That part of the emergence basis is implicit is particularly important for the claim that the bulk \textit{metric} emerges from entanglement. Besides entanglement, the necessary boundary data here include the boundary metric. Thus, if one by `spacetime' means metric (possibly plus manifold), then both the (alleged) emergent level and the underlying emergence basis include spacetime. The ``fabric of spacetime," as Swingle puts it, would then comprise of both entanglement and another spacetime, albeit one with an extended dimension less. Section \ref{DeterminationNovelty} therefore questions whether the novelty condition for emergence is satisfied when the bulk metric is claimed to emerge from boundary entanglement. By contrast, for the claim that gravitational dynamics emerges from entanglement, none of the implicit elements of the emergence basis is problematic from the perspective of novelty, as Section \ref{Gravitational} argues. In the case of the claim that topological connectivity emerges from entanglement, it is less clear what the additional implicit elements of the emergence basis are and, therefore, unclear whether novelty is satisfied. Section \ref{Topology} discusses some reasons, though, to be concerned that novelty might not be satisfied in this case either.

\section{The Hubeny-Rangamani-Takayanagi (HRT) Formula}\label{Sec:HRT}
\begin{figure}
	\begin{center}
		\includegraphics[scale=0.4]{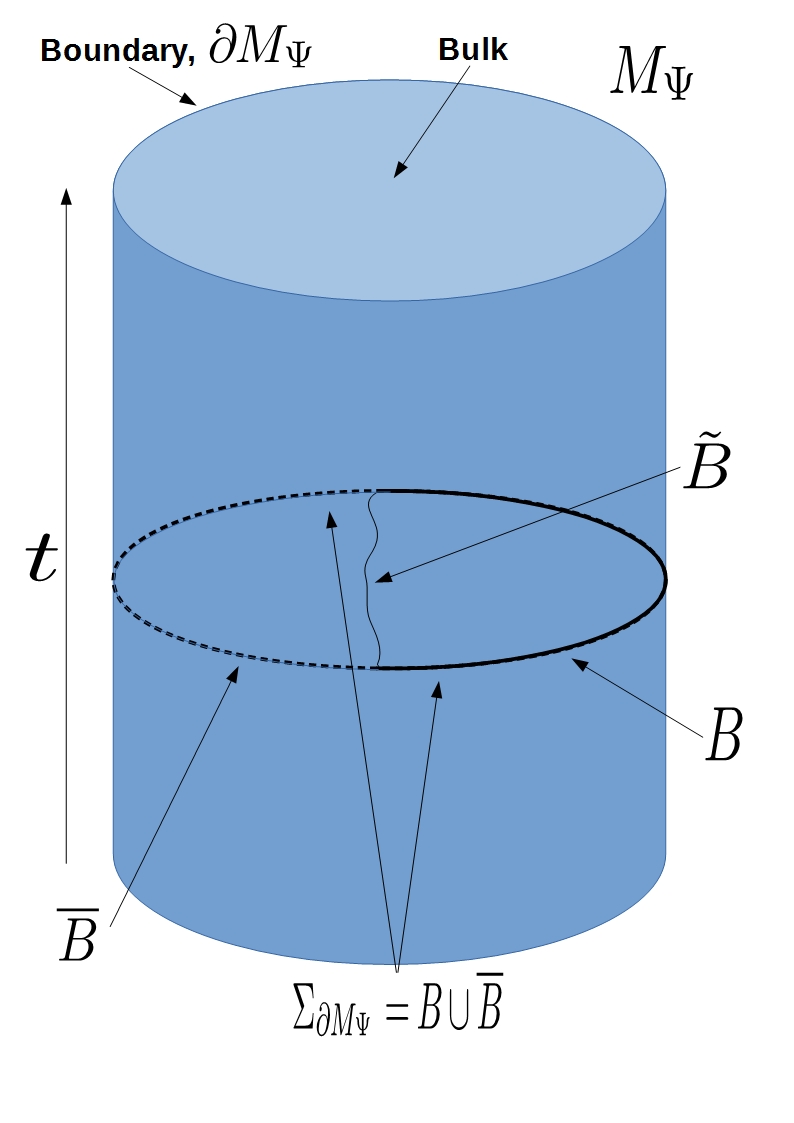}
		\caption{\label{cylinder} Illustration of the AdS/CFT correspondence. With an AdS metric in the interior, this is more precisely an illustration of the universal covering of global $AdS_3$. Illustration is taken from \citet[9676]{jaksland_entanglement_2021}, CC BY 4.0.}
	\end{center}
\end{figure}

The AdS/CFT correspondence is a (conjectured) duality in string theory. In being a duality, it conjectures that two apparently very different physical theories are formally isomorphic in such a way that they agree on all their empirical predictions.\footnote{See \citet{de_haro_schema_2018} for how best to theorize what dualities are.} In the AdS/CFT correspondence (in the form considered here), the two theories are: (AdS side) a theory with an asymptotically anti-de Sitter metric that couples to matter and is subject to equations of motion,\footnote{\label{AdS5xS5}In one concrete example of the correspondence, the AdS side consists of a type IIB string theory defined on $AdS_5 \times S^5$. In the limit where the string length is much smaller than the characteristic length scale of the spacetime background and the string coupling is much smaller than one, the AdS side is well approximated by semi-classical (super)gravity on $AdS_5$.}  i.e., a theory of gravity, and (CFT side) a conformal field theory defined on the asymptotic boundary of its AdS dual which, for typical choices of boundary conditions, entails that the metric on the CFT side is static,\footnote{In the concrete example of note 2, the CFT side consists of a $\mathcal{N}=4$ super Yang-Mills theory defined on a spacetime conformal to the 4-dimensional asymptotic boundary of $AdS_5$. The limit where the AdS side is approximated by semi-classical gravity corresponds, on the CFT side, to the limit where the gauge group goes to infinity and the t’Hooft coupling is large but finite.} i.e., it is a theory without gravity that has one extended\footnote{That this is one \textit{extended} dimension less is an important qualification since the AdS side, in the concrete example of note 2, is defined on a 10-dimensional spacetime, but the five of them are compactified (those of $S^5$).} dimension less than the AdS side. For the universal covering of global AdS, the CFT side can be defined on an Einstein static Universe. When the extended dimensions of the AdS side are $AdS_3$, this entails that the CFT can be defined on $S^1 \times \mathbb{R}$. This lends the correspondence to an illustration as a cylinder (see figure \ref{cylinder}), where the CFT side is defined on the boundary of the cylinder and the AdS side is defined on the interior of the cylinder (if remembering that this interior must have asymptotically AdS metric). 

The relation between CFT entanglement and features of the AdS geometry, in the form of the HRT formula, can then be introduced in this way \citep[following][]{rangamani_holographic_2017,jaksland_entanglement_2021}: Consider a CFT state $\ket{\Psi}$ whose AdS dual is given by a manifold equipped with a classical asymptotically AdS metric (and matter fields), for brevity denoted $M_{\Psi}$ here. Let $\partial M_\Psi$ be the asymptotic boundary of $M_{\Psi}$ and let $\Sigma_{\partial M_\Psi}$ be a Cauchy surface of $\partial M_{\Psi}$, i.e., $\Sigma_{\partial M_\Psi}$ defines an instant of time on the boundary whose domain of dependence\footnote{The domain of dependence of a surface is the set of all those points, $p$, on the manifold such that every inextendable causal curve containing $p$ intersects the surface in question.} is $\partial M_\Psi$.\footnote{This construction is possible because the AdS asymptotic boundary is globally hyperbolic even though the AdS spacetime is not.} The CFT degrees of freedom have support on $\Sigma_{\partial M_\Psi}$. We therefore have $\ket{\Psi} \in \mathcal{H}_{\Sigma}$, where $\mathcal{H}_{\Sigma}$ is the CFT Hilbert space. Defining a spatial subregion, $B$, of $\Sigma_{\partial M_{\Psi}}$ and its complement $\overline{B}$ (see figure \ref{cylinder}), the literature discussed below \textit{assumes} that this Hilbert space decomposes into a tensor product of Hilbert spaces associated with the degrees of freedom in $B$ and $\overline{B}$, respectively, i.e., $\mathcal{H}_{\Sigma} = \mathcal{H}_{B} \otimes \mathcal{H}_{\overline{B}}$.\footnote{In relativistic quantum field theories including in CFTs, the Hilbert space cannot, in general, be decomposed like this into a tensor product of Hilbert spaces of open regions, and traces, like those used below to evaluate the entanglement entropy, are not well-defined. This is because the algebras of observables with support in local open regions like $B$ and $\overline{B}$ are type III von Neumann algebras in relativistic quantum field theories \citep{yngvason_role_2005}. The account of the HRT formula in terms of Hilbert spaces and traces is, therefore, accompanied by the---sometimes only implicit---introduction of a UV cutoff at the scale $1/\epsilon$, in which case the construction is valid (assuming that the CFT is sufficiently simple and not, for instance, a gauge theory where an analog of the HRT formula requires further setup and is still tentative \citep{kamal_ryu-takayanagi_2019}). A consequence of this is that the entanglement entropy of a given boundary region depends on the cutoff and diverges for $\epsilon \rightarrow 0$ \citep{casini_lectures_2022}. On the AdS side, the CFT cutoff corresponds to the introduction of a regulated conformal boundary at $z=\epsilon$ in Fefferman-Graham coordinates. See \citet{taylor_renormalized_2016} for further details on the renormalization of the quantities in the HRT formula.} A state, $\ket{\Psi} \in \mathcal{H}_{\Sigma}$, is entangled with respect to the bipartitioning $\mathcal{H}_{B} \otimes \mathcal{H}_{\overline{B}}$ when $\ket{\Psi}$ cannot be written as a convex sum of products of pure states in, respectively, $\mathcal{H}_{B}$ and $\mathcal{H}_{\overline{B}}$. This entanglement can be quantified using the entanglement entropy: $S_B = -\tr ( \rho_B \log ( \rho_B ) )$ with the density matrix $\rho_B = \tr_{\overline{B}} ( \ket{\Psi} \bra{\Psi})$.

On the AdS side, consider first a case where the symmetries are such that the Cauchy surface on the boundary, $\Sigma_{\partial M_{\Psi}}$, can be uniquely extended into the bulk as a constant time slice, $\tilde{\Sigma}_{M_{\Psi}}$.\footnote{More precisely for this to be the case, the CFT state must respect a timelike Killing field of the boundary geometry \citep[46]{rangamani_holographic_2017}.} Let $\tilde{B}$ be the minimum area codimension-2 surface of $\tilde{\Sigma}_{M_{\Psi}}$ that is homologous\footnote{$\tilde{B}$ must be smoothly retractable to $B$.} to $B$ such that its asymptotic boundary (or endpoints) separates $B$ from $\overline{B}$, i.e. $\tilde{B} \subset \tilde{\Sigma}_{M_{\Psi}}$ and $\partial \tilde{B} = \tilde{B} \cap \partial B$ (see figure \ref{cylinder}).\footnote{\label{Footnote:BEmpty}Some subtleties arise if there are more distinct surfaces with minimum area. Notice, too, that when $\overline{B}=\emptyset$ and $\ket{\Psi}$ is pure, $\tilde{B}=\emptyset$.} In the limit where the AdS side is a classical theory of gravity, which on the CFT side corresponds to the large $N$ limit of the gauge group, the original Ryu-Takayanagi \citeyearpar{ryu_holographic_2006} formula then states that 
\begin{equation} \label{eq:HRT}
S_B = \frac{\mathrm{Area}(\tilde{B})}{4 G_N \hbar}
\end{equation}
where $\mathrm{Area}(\tilde{B})$ is the area of $\tilde{B}$.\footnote{When moving beyond the large $N$ limit by including $1/N$ corrections to $S_B$, the bulk becomes semi-classical and the right hand side of the HRT formula will include a subleading contribution from bulk entanglement. Apart from the subtleties mentioned in note 25, the literature discussed below assumes that these corrections vanish. The considered emergence claims are about this counterfactual situation and the evaluation of these claims will keep that in mind. With that said, there is no reason to expect that future results that take quantum corrections into consideration will change the present analysis. Whether there is entanglement, too, on the AdS side seems irrelevant for whether spacetime on the AdS side might emerge from (some of the) entanglement on the CFT side.}

In cases where there is no unique extension of $\Sigma_{\partial M_{\Psi}}$ into the bulk, one must consider the whole family of non-timelike codimension-1 surfaces whose boundary is $\Sigma_{\partial M_{\Psi}}$. On each of these, there will be a codimension-2 surface that satisfies the criteria above (remembering that it might be the empty set, see note 10). The HRT formula, in its reformulation by \citet{wall_maximin_2014}, then states that Eq. (\ref{eq:HRT}) still holds for the surface among these, $\tilde{B}$, with maximum area.\footnote{Strictly speaking, the covariant generalization of Eq. (\ref{eq:HRT}) by \citet{hubeny_covariant_2007} is equivalent to the maximin formalism introduced here if the bulk satisfies the null curvature condition, $R_{ab} k^a k^b \geq 0$ for null $k^a$ where $R_{ab}$ is the Ricci tensor, or if an appropriate stability condition is satisfied \citep{akers_quantum_2020}.} Wall appropriately refers to these as ``maximin surfaces," but they will be referred to below as ``extremal surfaces" to match the terminology with that of the literature under discussion.

The HRT formula is part of the general dictionary that exists between the AdS and CFT side, and this particular entry states that entanglement entropies on the CFT side are dual to particular areas on the AdS side. This shows that there, loosely speaking, is a relation between facts about the spacetime on the AdS side and the entanglement structure on the CFT side. Swingle, however, makes a much stronger claim. Reviewing the HRT formula and some results derived from it, he finds that it ``justifies the claim that spacetime arises from entanglement" \citep[354]{swingle_spacetime_2018}. Swingle is not alone in expressing the implications of the HRT formula in these stronger terms. Under the telling title ``Spacetime equals Entanglement," \citet[370]{nomura_spacetime_2016} refer to the HRT formula for the claim that ``[r]ecently it has become increasingly clear that quantum entanglement in holographic descriptions plays an important role in the emergence of the classical spacetime."\footnote{When \citet{nomura_spacetime_2016} say that entanglement only ``plays an important role," they do admit that elements other than entanglement could play a role as well. However, for the emergence claim to be true, these elements can hardly include another spacetime. More on this in Section \ref{DeterminationNovelty}.} \citet[3]{bao_towards_2019} even describe it as an ``expectation" in the literature ``that the bulk should emerge from the entanglement structure of the boundary state." \citet[1]{burda_holographic_2019} likewise finds that the HRT formula supports ``the idea that the bulk spacetime emerges from the entanglement structure of the boundary field theory;" \citet[2328]{van_raamsdonk_building_2010} speculates, based on the HRT formula, ``that the intrinsically quantum phenomenon of entanglement appears to be crucial for the emergence of classical spacetime geometry" \citep[see also][198]{van_raamsdonk_spacetime_2020}; and the HRT formula is among the results that, according to \citet[9]{bianchi_architecture_2014}, ``illustrates the role of quantum entanglement as a key mechanism underlying the emergence of spacetime geometry." 

\section{Emergence}\label{Emergence}
Emergence is, in philosophy, often associated with features that, from the perspective of an underlying level of reality or description, are completely inexplicable. Such ``radical emergence" is the antithesis of reduction and occurs when ``novel laws, properties and processes come from nowhere" \citep{bishop_contextual_2022}. However, \citet[855]{wayne_emergence_2009} notice that `emergence' is used rather differently in the context of quantum physics. Here, both philosophers and physicists use the term ``emergence" when describing how ``classical physics [...] \textit{results from} quantum processes, that is, that classical physics is reductively explainable in terms of quantum theory." On this usage of `emergence,' the emergent (classicality) comes from somewhere (quantum processes). It is a kind of (at least weak) reductive explanation whereby inexplicability is not the characteristic feature of emergence.\footnote{This is also the predominant understanding of `emergence' in nuclear physics \citep{luu_topic_2021}.} Rather, what makes particular (proposed) reductive explanations of classicality from quantum processes instances of emergence seems to be that the emergent is novel compared to the emergence basis. The usage of `emergence' among philosophers and physicists working on the appearance of classicality in quantum physics is, in other words, more similar to the understanding of emergence proposed by \citet{butterfield_emergence_2011,butterfield_less_2011} who explicitly argues that emergence is compatible with reduction. 

Building on Butterfield's proposal, \citet[7285]{crowther_as_2021} has explicated the ``conception of emergence" that she finds "is now familiar in the philosophy of physics generally, and the philosophy of QG [quantum gravity] in particular." On this conception, emergence is a relation between an emergent and its emergence basis where the relation is such that the emergent ``is at once \textit{dependent} on," ``\textit{novel} compared to," and ``\textit{autonomous} from" the emergence basis \citep[7284, emphasis in original]{crowther_as_2021}. As already indicated, emergence-talk is frequent too among physicists working on quantum gravity, but neither Crowther nor anyone else seems to have analyzed whether the usage of `emergence' among these physicists generally aligns with the understanding that Crowther finds predominant among philosophers of quantum gravity. This does raise a methodological problem for a project like this that ultimately seeks to assess these physicists' emergence claims: What understanding of emergence should this assessment be based on? When Crowther's analysis is used here, it is for two reasons. First, with the emergence of spacetime in quantum gravity being a particular example of the ``emergence" of classicality from quantum physics, it seems reasonable to expect that physicists also here use `emergence' for a kind of reductive explanation, i.e., they could well be using `emergence' in the sense explicated by Crowther. Second, as Crowther states, her explication of emergence is predominant among philosophers of quantum gravity who are, among other things, trying to draw the metaphysical implications of results like these that are summarized as showing that spacetime emerges from entanglement. These philosophers will be prone to understand the physicists' emergence claims following Crowther's explication and assessing whether Crowther's conditions for emergence are satisfied by these claims is therefore at least important for what metaphysical picture philosophers of quantum gravity should derive from these results. This leaves it open that the usage of `emergence' among physicists complies with another internal convention. It is worth noticing, though, that Crowther's explication has the advantage of maintaining the distinctiveness of emergence from reduction and a certain conceptual continuity with other uses of `emergence' through its emphasis on novelty while being weaker than radical emergence. Thus, one might argue that also the physicists \textit{should} abide by Crowther's explication of emergence, if they do not already do so. 

In the assessment below of the claims from the literature that spacetime emerges from entanglement, the focus will be on Crowther's novelty condition and an aspect of the dependence condition. Thus, the discussion will concern whether certain necessary conditions for emergence (in Crowther's sense) are met, but without an assessment of autonomy and the full dependence condition, no conclusion can be made regarding whether these claims do, in the end, qualify as emergence claims by Crowther's standards. 

Minimally, to meet the dependence condition, the relation between the emergent and the emergence basis must be appropriately asymmetric. As already mentioned above and as also \citet[7289]{crowther_as_2021} points out, this \textit{asymmetry condition} appears to be in conflict with the standard setup of the AdS/CFT correspondence and, therefore, with the claim of Swingle and others that spacetime emerges from entanglement within this framework. When it is nonetheless relevant to investigate whether these claims meet some of the other necessary conditions for emergence, it is because there might be ways to modify the mathematics or metaphysics of the AdS/CFT correspondence such that it would satisfy the asymmetry condition, a possibility that \citet[7290]{crowther_as_2021} explicitly mentions following work by \citet{de_haro_dualities_2017}. Furthermore, \citet[399]{crowther_spacetime_2022} considers whether the results of quantum gravity research calls for abandoning the asymmetry condition for emergence. 

The other component of the dependence condition concerns the kind of connection that must obtain between the emergent and the emergence basis. Not every two sets of elements that are asymmetrically related as well as novel and autonomous from each other are instances of emergence. This would make emergence too inclusive. The dependence condition, \citet[386]{crowther_spacetime_2022} explains, requires that the emergent is ``derivable from [...] and/or supervenient upon" the emergence basis. Supervenience is a necessary condition for derivability so Crowther's "and/or" should be read as requiring that the emergent is \textit{at least} supervenient on the emergence basis which agrees with Palacios' \citeyearpar[40]{palacios_emergence_2022} explication of this kind of emergence and the view that \citet[431]{crowther_decoupling_2015} herself expresses elsewhere.\footnote{\citet[section 5.2.2]{butterfield_emergence_2011}, who otherwise defends a very similar conception of emergence, argues that supervenience is not necessary for emergence.} The connection between the emergent and the emergence basis must, in other words, be such that there can be no difference in the emergent without a difference in the emergence basis. Conversely, the emergence basis must uniquely determine the emergent. This necessary \textit{determination condition} for emergence is counterbalance to the novelty condition. The determination condition ensures that one cannot just remove elements from the emergence basis until novelty is satisfied, and the novelty condition ensures that one cannot just add the emergent to the emergence basis to satisfy determination. This dynamic is exactly what will play out below in the investigation of the claims that spacetime (the emergent) emerges from entanglement (the emergence basis). 

The novelty condition \citet{crowther_as_2021} explicates as the necessary condition that the emergent is ``qualitatively different from" the emergence basis. If the emergent is just more of the same of some part of the emergence basis, then the novelty condition is not satisfied, and the emergent does not stand in the emergence relation to the emergence basis. Again, it is important to emphasize that this understanding of emergence ``does not require novelty to be the failure of reduction, deduction, explanation, or derivation" \citep[7285]{crowther_as_2021}. What it does require instead, though, is not easily formalized. As already \citet[1088]{butterfield_emergence_2011} observes, novelty ``is liable to be fixed contextually, and even to be vague or subjective." 

To illustrate this point and how, in general, the determination condition and the novelty condition will be used, consider the claim that the mind emerges from the brain. For this claim, determination might fail if the extended mind hypothesis proves to be true since more would be needed in the emergence basis---something apart from the brain---to determine the mind. Novelty, in turn, might fail if panpsychism proves to be true since mind would then, in some sense, already be present in the emergence basis. The qualification `in some sense' is illustrative of the fact that the judgment whether the novelty condition is satisfied depends on an often non-trivial assessment of sameness. If our original claim concerned mind in the form of human-like consciousness, then this could well be argued to be ``qualitatively different," as Crowther requires of novelty, from the kind of mind that pervades everything according to some versions of panpsychism. Accordingly, one might therefore find that the novelty condition is after all satisfied. Such questions about sameness will likewise occur in the discussions below.

Crowther's own assessments of novelty in other cases of (alleged) spacetime emergence can give some direction for these discussions. One is in the context of the quantum gravity approach known as causal set theory. Here ``spacetime" is proposed to emerge from a so-called causal set which is a graph structure consisting of events connected by a partial order relation.\footnote{It remains a conjecture that spacetime is, in fact, determined by this structure provided by causal set theory.} About novelty, \citet[7292]{crowther_spacetime_2022} remarks that ``causal sets differ remarkably from spacetime" and goes on to give the example that lengths and duration are conjectured to be derivable from the causal set while ``there is nothing on the fundamental level corresponding to lengths and durations" \citep[7292]{crowther_spacetime_2022}.\footnote{For novelty and spacetime emergence in loop quantum gravity, see \citet[7292]{crowther_spacetime_2022}; same for string theory, see \citet[chapter 9]{huggett_out_2025}.} How exactly this assessment is made remains a little vague: What would length and duration look like in a formal representation like the causal set? Plausibly, Crowther's "lengths and durations" is a pedagogical way of referring to the metric tensor.\footnote{See \citet{jaksland_non-identity_2024} for a further discussion of the non-trivial relation that lengths and durations have to the causal set including what length and duration might mean in the first place.} If this is the case, then we can understand the novelty as involving the determination of this mathematical object from a basis that does not include such a mathematical object.

\section{Metric from Entanglement}\label{Metric}

Emergence is not the only concept that needs explication. As \citet{jaksland_many_2023} have pointed out about discussions of spacetime emergence in general, people already disagree about what spacetime is in general relativity. This entails that, in discussions like these about spacetime emergence, it can often be ambiguous if not outright unclear what these discussions are about when they are had in terms of `spacetime.' Furthermore, as Jaksland and Salimkhani also point out, it can often be very difficult to make any progress in such discussions until the `spacetime'-talk is replaced with references to the specific spatiotemporal aspects one is interested in. Crowther's discussion about spacetime emergence in causal set theory is a case in point where the actual assessment required the identification of a spatiotemporal aspect, the metric, before the emergence claim could be assessed. Likewise, consider the question how much of bulk spacetime can be recovered from entanglement entropies using the HRT formula. This is not immediately a question we can pose to the formalism. What we can ask is whether the bulk metric, signature, dynamics, topology, etc. can be recovered from entanglement entropies using the HRT formula. 

Among these, the bulk metric is the spatiotemporal aspect that is most immediately associated with the areas of codimension-2 surfaces that the HRT formula relates to boundary entanglement entropies. These bulk areas are given by the area functional which can be expressed in terms of the bulk metric, $g_{\mu\nu}$, and an embedding function, $X^\mu$, from the codimension-2 surface to the bulk manifold:
\begin{equation}\label{Area}
    \mathrm{Area}(g,X)=\int_{\tilde{B}} d^{d-1} \sigma \sqrt{\mathrm{det} \left( \frac{\partial X^\mu}{\partial \sigma^a} \frac{\partial X^\nu}{\partial \sigma^b} g_{\mu\nu} \right) }
\end{equation}
where $\partial_a X^\mu \partial_b X^\nu g_{\mu\nu}$ is the induced metric on the codimension-2 surface. In discussions of the emergence of spacetime from entanglement in the physics literature, the metric is indeed often the specific spatiotemporal aspect under discussion. And \citet[81]{ney_quantum_2021}, in the most recent philosophical discussion of the HRT formula, also focuses on the metric and frames the discussion as the question of whether ``facts about the AdS metric" or ``facts about the entanglement entropy" are more fundamental. However, one important point of this paper is that this identification of spacetime with metric (plus manifold) is not unanimous in the physics literature. Others who discuss their results in terms of the emergence of spacetime from entanglement are concerned with different spatiotemporal aspects as Section \ref{Gravitational} and \ref{Topology} will show. In this and the next section, however, the focus will be on whether the HRT formula can warrant the claim that the bulk metric emerges from entanglement.

That it is even a possibility that the HRT formula might support the stronger conclusion that the bulk metric emerges from entanglement is because there can be many instantiations of the HRT formula for a given AdS/CFT dual. Thus, many areas on the AdS side can be related to entanglement entropies in the CFT side whereby we would have many instances of Eq. (\ref{Area}) with different areas on the left hand side but the same metric on the right hand side. Could this be sufficient to determine that metric?

One immediate worry is that these areas are areas of bulk extremal surfaces that, by construction, are anchored on a Cauchy surface (an instant time slice) on the boundary and are spacelike everywhere in the bulk. Thus, what is immediately related to entanglement entropies via the HRT formula is only something spatial, in this sense, irrespective how many instances of them one might have. Especially in the context of asymptotically AdS spacetime this can be an important distinction to make. AdS spacetimes are not in general globally hyperbolic. There is generally no Cauchy surface in AdS spacetime, i.e., no spacelike surface that is at most intersected once by any timelike curve and whose domain of dependence is the whole manifold. This entails that there is no spacelike surface where the initial data on this surface together with the dynamics determine the rest of the spacetime. The Cauchy problem is not well-posed in spacetimes that are \textit{not} globally hyperbolic. This raises the concern that the HRT formula cannot be used to recover the full spacetime from entanglement irrespective of how many spatial facts one can recover from it. 

In the context of AdS/CFT, however, this problem goes away. The Cauchy problem is well-posed in AdS spacetime if appropriate boundary conditions are imposed at the full timelike boundary at spatial infinity (which in figure \ref{cylinder} is the entire exterior of the cylinder) \citep{bantilan_cauchy_2021}. And fixing these boundary conditions is necessary for the setup of the AdS/CFT correspondence. Most evidently, with Dirichlet boundary conditions which is the most typical in AdS/CFT \citep[footnote 48]{hubeny_ads/cft_2015}, one of the boundary conditions necessary for a well-posed Cauchy problem includes a choice of induced metric on the boundary, and this induced metric also serves as as the metric for the spacetime on which the CFT is defined \citep{marolf_conserved_2014}.\footnote{For a bulk scalar field, $\phi$, Dirichlet boundary conditions imply the typical constraint of AdS/CFT that $\phi = z^{d/2-\sqrt{d^2/4+m^2}}$ near the boundary at $z=0$ (in Poincaré AdS coordinates) \citep{minces_scalar_2000}. For further details on how reflective Dirichlet boundary conditions allow for a well-posed Cauchy problem, see \citep{bantilan_cauchy_2021}. See \citet{compere_setting_2008} for how to do AdS/CFT with von Neumann and mixed boundary conditions which actually allow the metric to be a dynamical field on the CFT side.} Thus, the CFT remains ill-defined without such a choice of boundary conditions, and different choices of boundary conditions (including the type of boundary conditions) correspond to different versions of AdS/CFT. That the AdS/CFT correspondence requires that the Cauchy problem is well-posed in the bulk is, of course, hardly surprising because the CFT side, to be a consistent quantum field theory, must admit a Cauchy evolution (using unitary evolution of the CFT state). So to preserve the duality, the same must be the case on the AdS side, and the Cauchy problem must therefore be well-posed also in the bulk.\footnote{Notice that this still entails that causality requirements in the bulk can be much weaker than global hyperbolicity. While closed timelike curves are ruled out, bulk spacetimes can be arbitrarily close to having closed timelike curves and still have a sensible CFT dual \citep{hubeny_causally_2005}.} Assuming AdS/CFT, we can, in other words, be hopeful that the full bulk spacetime can be recovered using the HRT formula despite all its extremal surfaces being spacelike. We can be hopeful that the areas of the extremal surfaces that the HRT formula relates to CFT entanglement entropies uniquely determine the metric in the AdS bulk. 

Whether this can be brought to fruition is precisely the problem that Bao et al. \citeyearpar{bao_towards_2019,bao_more_2021} try to answer, as detailed below, and the way they frame this problem is very illustrative of the point of this paper. There are, \citet[3]{bao_towards_2019} explain, two questions one might ask: ``which CFT states correspond to a dual bulk geometry, and how does this geometry emerge from boundary degrees of freedom? Here we endeavor to answer this second question: how, precisely, does the bulk spacetime arise from the boundary?" \citep[3]{bao_towards_2019}. The HRT formula is the clue for how to answer because, as they state earlier, ``[t]he Ryu–Takayanagi and Hubeny–Rangamani–Takayanagi formulae suggest that bulk geometry emerges from the entanglement structure of the boundary theory." More precisely, the HRT formula relates entanglement entropies to the areas of boundary-anchored bulk extremal surfaces ``so the HRT formula [...] naturally leads to a purely geometric question: do the areas of extremal surfaces anchored to the boundary of a manifold uniquely determine its geometry" \citep[4]{bao_towards_2019}. And while this ``problem of determining a bulk metric from the areas of all possible boundary-anchored extremal surfaces is very overconstrained, and the holography community often implicitly assumes that such a result must be true" \citep[4]{bao_towards_2019}, actually showing it is true is challenging. 

Before turning to why this is challenging, notice the two terminological shifts that occur here. What is initially framed in terms of ``spacetime" and ``geometry" turns out to be a question about the bulk metric, as will become even clearer below. Second, questions about how spacetime might ``emerge" or ``arise" from the boundary transform into a question about \textit{determining} the bulk metric. This is a rather strong indication that \citet{bao_towards_2019} do not see a contradiction in terms between emergence and determination and that they are rather endorsing determination as a necessary condition for emergence. 

About the question \citet{bao_towards_2019} are raising, one might expect that writing up instances of the area functional, Eq. (\ref{Area}), for all the possible values of areas of boundary-anchored extremal surfaces would determine the metric. However, proceeding like this is not easy. First, from the entanglement entropies, $S_B$, only the areas of the extremal surfaces, but not the surfaces themselves, are known via the HRT formula. Thus, besides the metric, the surface being integrated over, $\tilde{B}$, and therefore also the embedding function, $X^a$, are unknown and, importantly, vary for every instance of Eq. (\ref{Area}). What we do know from the construction of the HRT formula is that the embedding is such that it extremizes Eq. (\ref{Area}) given the (unknown) metric, $g_{\mu\nu}$, and that this resulting extremal surface, $\tilde{B}$, must be anchored to $B$ at the boundary. But when starting from the area functional, these conditions must be imposed separately. Furthermore, determining the metric is complicated because, in Eq. (\ref{Area}), the known area is obtained from integrating over the metric. 

Therefore, Bao et al. \citeyearpar{bao_towards_2019,bao_more_2021}, who offer the hitherto most detailed prove that the bulk metric is uniquely determined by boundary data, take an approach where the metric is solved for in the Jacobi equation. The Jacobi equation is a generalization of the geodesic deviation equation. Remember from general reletivity that geodesics are extremal 1-dimensional surfaces, i.e., curves that are (local) extrema of the \textit{length} functional. One can consider a continuous one-parameter family of closely spaced geodesics. The partial derivative of the geodesic (at each point along it) with respect to this parameter gives the so-called deviation vector, a vector from each point along that geodesic to points along the geodesic infinitesimally close to it. The geodesic deviation equation relates this deviation vector---more precisely its components normal to the geodesic---to the metric via the Riemann tensor. The equation can be interpreted as capturing how a geodesic varies in response to a small variations of its boundary conditions \citep[6-7]{engelhardt_surface_2019}. Intuitively, this must depend on the metric because the metric determines what curves are (local) extrema of the length functional, i.e., geodesics, and, therefore in particular, which of the curves with a given set of boundary conditions that is a geodesic. How a geodesic varies when its boundary conditions are varied must therefore depend on the metric (here via the Riemann tensor). For the same reasons, however, knowing how geodesics vary in response to changes in the boundary conditions can be used to extract information about the metric. 

Similarly, $\tilde{B}$ is an extremum of the area functional, and we can form a one-parameter family of such closely spaced, boundary-anchored, extremal codimension-2 surfaces. For this family, one can likewise define deviation vectors that point from one of these codimension-2 surfaces to the one that is infinitesimally close. These deviation vectors must also satisfy certain conditions that depend on the metric and these are captured by the Jacobi equation. More precisely, the Jacobi equation relates the normal components of the deviation vector of a continuous family of extremal surfaces to the induced metric and extrinsic curvature of those surfaces, to the metric, and to the Riemann tensor. In compact form, the Jacobi equation states that the Jacobi operator, which can be expressed in terms of the aforementioned elements, acting on the deviation vectors vanishes. Again, the interpretation of the equation is that it captures how the extremal surface must vary in response to a small variations of its boundary conditions. Thus, the Jacobi operator encodes the facts about the metric that determine, for a given extremal surface, what extremal surface is infinitesimally close to it as captured by its deviation vector. 

As was the case for the area functional, the problem is again that neither the extremal surfaces nor the bulk metric is known. However, from the HRT formula, we know the area of both the perturbed and unperturbed extremal surface. Using this information, Bao et al. \citeyearpar{bao_towards_2019,bao_more_2021} succeed in solving this partial differential equation for the metric (which is very non-trivial), and the way they ultimately summarize the results seems promising for the proposal that the metric emerges from entanglement: ``we conclude that the bulk metric in $\mathcal{R}$ [the spacetime neighborhood of $\tilde{B}$] is fixed by boundary entanglement entropies" \citep[3]{bao_more_2021}. This is their answer to the question they, as already quoted, variously states as ``how, precisely, does the bulk spacetime \textit{arise} from the boundary" and ``how does this geometry \textit{emerge} from boundary degrees of freedom." In the philosophical literature, this conclusion in taken up by Ney who, with direct reference to \citet{bao_towards_2019}, also reports---without going into the details of the derivation or the interpretation of its conclusions---``that entanglement entropies on the boundary are sufficient to uniquely fix the metric (up to diffeomorphisms) everywhere in the neighborhood of the extremal surfaces" \citep[84]{ney_quantum_2021}.\footnote{As \citet{ney_quantum_2021} also notices, this result only determines the bulk metric locally in the neighborhood of every possibly extremal surface. Thus, if there are regions of bulk spacetime that are not reached by such surfaces, then the metric of those regions will not be determined by boundary data. It remains an open question, therefore, whether the full bulk metric is determined by boundary data, but we will not discuss this problem further here.}

Looking closer at the derivation, however, more elements from the boundary go into the determination of the bulk metric, as Bao et al. (\citeyear[6]{bao_towards_2019}; \citeyear[3]{bao_more_2021}) readily recognize. Besides the entanglement entropies, the necessary boundary data include the induced manifold, induced metric, and induced extrinsic curvature on the boundary as well as the embedding of $B$ into the boundary manifold. This is not too surprising considering that the Jacobi equation governs how the extremal surfaces vary in response to perturbations of their boundary conditions, i.e., how $\delta \tilde{B}$ relates to $\delta B$. To have any hope of determining the metric with the Jacobi equations, we must at least know what the boundary conditions are \textit{and} what boundary conditions are associated with what extremal surfaces, where the latter also serves to ensure that, instead of the HRT fomula just giving us a list of areas corresponding to CFT entanglement entropies, the list is labeled with the boundary conditions of the unknown surface that each area belongs to. Knowing this amounts to knowing $B$ and $\delta B$ which amounts to knowing the embedding from these to the boundary manifold and knowing the boundary induced metric. Furthermore, we must know the embedding from this intrinsic geometry to the bulk geometry, which is what is captured by the extrinsic curvature.

\section{Determination and Novelty}\label{DeterminationNovelty}
Bao et al. ask ``how, precisely, does the bulk spacetime \textit{arise} from the boundary," notice that there is ``the expectation that the bulk should emerge from the entanglement structure of the boundary state," and give as an answer ``that the bulk metric in $\mathcal{R}$ is fixed by boundary entanglement entropies." Spacetime, in the meaning metric, emerges from boundary entanglement entropies, or so they seem to suggest. However, that the boundary data needed to determine the bulk metric include more than entanglement entropies immediately raises concerns about whether the determination condition is satisfied. The emergent, in this case, is the bulk metric, and the emergence basis is claimed to be entanglement (entropies). And while the entanglement entropies are part of the boundary data necessary to determine the bulk metric, this boundary data also include several other elements. The emergence basis, to comply with the determination condition, should therefore include these additional elements as well. To satisfy the determination condition, the proposal must be that the bulk metric emerges from the induced metric and intrinsic curvature on the boundary as well as the entanglement entropies for known boundary regions. Since Bao et al. are rather clear that this other boundary data are necessary to determine the bulk metric, it is arguably rather puzzling why they, in their conclusion, nevertheless only mention entanglement entropies. The explanation cannot be that they do not intend this to be an emergence claim because fixing or determining the metric would entail precisely the same requirement. However, the likely explanation for this omission is not philosophical carelessness either. 

Compare the claim that spacetime emerges from or is determined by entanglement to saying that the thermodynamic property temperature emerges or is determined by the average kinetic energy in statistical mechanics.\footnote{The relation between temperature and average kinetic energy is often considered a paradigmatic example of reduction, so some might object that their relation cannot be that of emergence. Again, however, `emergence' is here used in the sense typical in physics and philosophy of physics where emergence is compatible with reduction. Indeed, \citet[389]{crowther_spacetime_2022} uses temperature to illustrate this understanding of `emergence'.} Here, the determination condition likewise appears not to be satisfied. As \citet{bishop_contextual_2006} point out, the concept of temperature only makes sense under the assumption of thermodynamic equilibrium (besides even further assumption). To satisfy the determination condition, these additional assumptions must be added to the emergence basis. Nevertheless, there is arguably still a sense in which it would be correct to say that temperature emerges from or is determined by average kinetic energy. The purpose might be to differentiate the view of Maxwell from that of, for instance, Herapath who, in work on the kinetic theory of gas from the 1820s, proposed that temperature is determined by the average momentum \citep{truesdell_early_1975}. Such a \textit{contrasting emergence claim} is made in a context where certain elements of the emergence basis, the thermodynamic equilibrium in this particular case, are implied by the context. For contrasting emergence claims, the determination condition can be implicitly satisfied, in the sense that once these implicit elements of the emergence basis are included, the determination condition is explicitly satisfied.

As argued below, all the various statements suggesting that spacetime emerges from entanglement in the quantum gravity literature should charitably be understood as contrasting emergence claims. In the case of Bao et al., their whole derivation takes place in the context of the AdS/CFT correspondence. Here, the boundary manifold and the induced metric on the boundary must be assumed already when defining the quantum field theory on the CFT side since these serve as the background spacetime of the CFT and serve to define the necessary boundary conditions for the AdS side. Many bulk metrics will be compatible with these boundary conditions, and their CFT duals will therefore all be defined on the same spacetime background. What Bao et al. show is that the entanglement entropies associated with boundary regions like $B$ are the only additional elements one needs to know about the CFT side to determine the bulk metric (in the region $\mathcal{R}$). Entanglement entropy is the state-specific element on the CFT side that must be added to the generic setup of AdS/CFT to determine the metric in the regions foliated by boundary-anchored extremal surfaces. Given this contrast class, entanglement is the element of the emergence basis worth highlighting just like average kinetic energy in the thermodynamic context. That the emergence basis contains more than entanglement is made more explicit in some of the other quotation from Section \ref{Sec:HRT}. They contained the qualifications that entanglement ``plays an important role" \citep[370]{nomura_spacetime_2016}, that it ``appears to be crucial" \citep[2328]{van_raamsdonk_building_2010}, and that it is ``a key mechanism" \citep[9]{bianchi_architecture_2014} for the emergence of spacetime which can be read as signaling that these are contrasting emergence claims where entanglement is merely emphasized because it, given the context, is what is worth highlighting from the emergence basis.

While the claims in the literature that spacetime emerges from entanglement can thereby be seen as satisfying the determination condition once they are interpreted as contrasting emergence claims, leaving some of the emergence basis implicit is not without consequence. In the case where one by `spacetime' means metric, one of the implicit elements of the emergence basis is another metric. The contrasting emergence claim obscures that a metric, in the form of the bulk metric, is claimed to emerge from an emergence basis that includes another metric, namely the induced metric on the boundary. This must make us question whether the emergent (the bulk metric) is sufficiently qualitatively different from this emergence basis as required by the novelty condition because the emergence basis, too, includes a metric. After all, the emergent and the emergence basis now contains the same kind of mathematical object. 

As was the case for human-like consciousness and the kind of mind that pervades everything according to panpsychism, one might object that the bulk metric and the induced metric can still be qualitatively different. In the case of Bao et al. \citeyearpar{bao_towards_2019,bao_more_2021}, the bulk metric is the metric of a 4-dimensional manifold while the induced metric is the metric of a 3-dimensional manifold. The bulk metric couples to matter and is subject to equations of motion (the Einstein Field Equations) whereas this is not the case for the induced metric (more on this in Section \ref{Gravitational}).\footnote{Assuming Dirichlet boundary conditions.} One might think that the bulk metric and the induced metric are, at least, the same on the boundary, but even this is not the case. The induced metric on the boundary, which serves as the background for the CFT, can be expressed in terms of the bulk metric, but this expression will also depend on the embedding from the boundary to the bulk manifold. This, in turn, entails that the induced metric on the boundary does not even determine the bulk metric at the boundary. Only the induced metric together with the area variations do so \citep[appendix B]{bao_towards_2019}. Thus, there is a sense in which the whole bulk metric (and not just that away from the boundary) is different from the induced metric on the boundary. Whether this difference is sufficient to satisfy novelty is perhaps, as Butterfield says, ultimately ``vague or subjective." What is interesting, though, is that this is not the only case in quantum gravity where the question arises whether the determination of one metric from another can satisfy novelty and, therefore, candidate as an instance of emergence. In the broader context of superstring theory, for instance, one can find the suggestion that phenomenal spacetime emerges from target space, which implies that a 4-dimensional metric emerges from a 10-dimensional metric (though \citet{huggett_out_2025} and \citet{matsubara_spacetime_2018} warn against the realism with respect to target space that this implies). 

Though interesting, the question whether the determination of one metric from another can qualify as emergence will be set aside here. Because, irrespective of whether the emergence claim of Bao et al. might be vindicated like this, for the question of whether spacetime, in the meaning metric, emerges from entanglement, it is arguably an important difference whether spacetime, in this sense, emerges from entanglement or whether it emerges from entanglement and another spacetime. From the perspective of metaphysics, the former but not the latter raises questions about empirical coherence, about certain forms of naturalism \citep{le_bihan_priority_2018}, and about the fate of spacetime-based metaphysical frameworks such as Humeanism \citep{matarese_loop_2019,wuthrich_when_2020} and (spatiotemporal) mereology \citep{baron_curious_2020}. The lesson, in other words, is that metaphysicians must be cautious of scientists' use of contrasting emergence claims. Contrasting emergence claims can conceal that the novelty condition is not satisfied, i.e., that the emergent and an element of the emergence basis are not sufficiently qualitatively different. Contrasting emergence claims can therefore come across as more remarkable than they actually are. In particular, the contrasting emergence claim that spacetime, in the meaning metric, emerges from entanglement conceals that a metric features in both the emergent and the emergence basis. What the results show are \textit{not} that a metric can be determined from some pre-metrical entanglement structure, as one could be lead to think. Rather, the results show how the bulk data contained in the bulk metric are encoded on the boundary, and this data are distributed between several elements including boundary entanglement but also the induced metric on the boundary, i.e., this bulk geometric data are also partly encoded in boundary geometric data. This seems far from any alleged disappearance or non-fundamentality of spacetime one might glance from how these results are presented. 

\section{Gravitational Dynamics from Entanglement}\label{Gravitational}
Bao et al. \citeyearpar{bao_towards_2019,bao_more_2021} rather clearly intend their remarks about the emergence of spacetime and geometry to concern the bulk metric. When \citet[1]{burda_holographic_2019}, as quoted above, propose that ``spacetime emerges from the entanglement structure of the boundary field theory," the context also clearly indicates that `spacetime' here means metric. Others who discuss the emergence of spacetime from entanglement are less explicit about what they mean by `spacetime' and `geometry,' but this can still typically be inferred from the context. By `spacetime,' \citet{nomura_spacetime_2016}, for instance, seem to mean a triple consisting of a classical metric, a manifold, and quantum fields, though they never mention the words `metric' or `manifold.' What is interesting and important for those interested in the philosophical implications of these results is that, while what is meant by `spacetime' can typically be gleaned from the context, `spacetime' is not always just another word for the metric. Swingle, still writing under the title ``Spacetime from Entanglement," reviews a result developed in \citet{lashkari_gravitational_2014}, \citet{faulkner_gravitation_2014}, and \citet{swingle_universality_2014} and finds that it ``further justifies the claim that spacetime arises from entanglement" \citep[354]{swingle_spacetime_2018}. What is interesting is that \citet{faulkner_gravitation_2014} themselves describe the result as utilizing ``a quantitative connection between CFT entanglement and the dual spacetime geometry\footnote{This ``quantitative connection" is the HRT formula.} [...] to understand the emergence of \textit{spacetime dynamics} (i.e. gravity) from the CFT physics" \citep[2, emphasis added]{faulkner_gravitation_2014}. The result, as described in more detail below, is not concerned with determining the bulk metric from entanglement. Rather, it investigates what constraint the bulk (linearized) Einstein equations translates to on the CFT side, and the finding is that this is a constraint on entanglement entropy.

One might argue that also discussions of gravitational or ``spacetime dynamics" are still ultimately about the metric. After all, a metric is what solves the Einstein equations (when an energy-momentum tensor is given). Thus, when the Einstein equations in the bulk correspond to a constraint on entanglement entropy on the CFT side, this could be taken for another result that shows how entanglement determines the bulk metric in support of the claim that the metric emerges from entanglement. This would, however, be too quick as evidenced by the results that Swingle reviews. For any given metric, one can split it into a background metric and a perturbation term. The linearized Einstein equations then describe the first order perturbation of the Einstein equations only in terms of the perturbation term. These equations, in other words, describe the dynamically admissible metric perturbations away from the background metric to leading order. If the background metric is asymptotically AdS, then the linearized Einstein equations are equivalent to imposing everywhere in the bulk a local constraint on the relation between the area variation of boundary-anchored codimension-2 surfaces and the variation of their linearized holographic stress-energy tensor where the latter can be expressed as an integral over the asymptotic boundary of the same codimensions-2 surface. The codimension-2 surfaces can be chosen such that they are extremal, and their area is thereby dual, via the HRT formula, to the entanglement entropy of the CFT region corresponding to the asymptotic boundary of the codimension-2 surface. The linearized holographic stress-energy is in turn dual to what is known as the ``hyperbolic energy" of the same CFT region \citep[2]{lashkari_gravitational_2014}, which can be expressed an integral of the CFT energy density operator over that CFT region. Requiring that the entanglement entropy variation is equal to the variation in hyperbolic energy for all such CFT regions thus proves to be equivalent to imposing the linearized Einstein equations in the bulk \citep{lashkari_gravitational_2014,faulkner_gravitation_2014,swingle_universality_2014}.\footnote{\label{footnote:HRTmattercoupling}Notice that the derivation in all three references is run in the reverse, i.e., from the CFT constraint to the the linearized Einstein equations and that only the source-free linearized Einstein equations are obtained from the HRT formula of Eq. \ref{eq:HRT}. The linearized Einstein equations coupled to matter can be obtained by including quantum corrections to the HRT formula \citep{swingle_universality_2014}.} This CFT constraint is known as the first law of entanglement entropy, and the CFT states that satisfy this dynamical constraint have an AdS dual that satisfy the corresponding linearized Einstein equations \citep[see][for a non-technical review of this result and a discussion of some of its other philosophical implications]{jaksland_holography_2020}.\footnote{\citet{faulkner_nonlinear_2017} have extended this result to second order perturbations, but this is of no consequence for the conceptual point made here.} 

For present purposes, what is worth noticing about this result is that the background metric must be given. The result is not determining the bulk metric from boundary data. Rather, it shows that if the bulk metric is assumed to be AdS, then it is a constraint on entanglement entropy that ensures that the Einstein equations are satisfied in the bulk. This result is, in other words, an answer to the puzzle how there can be gravity in the form of a metric coupled to matter on the AdS side when there is no gravity on the CFT side. The answer is, at least to leading order perturbations away from an asymptically AdS metric, that this gravitational dynamics is encoded in the entanglement structure on the CFT side such that a particular constraint on the entanglement dynamics captures the dynamics otherwise encoded in the linearized Einstein equations. Spacetime, in the form of a metric, is present at both the AdS side and the CFT side, but it is only dynamical on the AdS side.\footnote{This echoes the distinction of \citet{linnemann_hints_2018} who, in the general context of emergent gravity, remark that there is a difference between ``the emergence of both the metric field and its dynamics (rather than just the dynamics)" \citep[2]{linnemann_hints_2018}. They do not, however, go into details about why they draw this distinction. The discussion here, however, is a way of cashing out this difference that is relevant too for emergent gravity in general.} 

What Swingle discusses in terms of how ``spacetime arises from entanglement" concerns the emergence of gravitational dynamics from entanglement and not the emergence of the metric. That there are such instances where claims about spacetime emergence are about the emergence of gravitational dynamics is, however, not surprising because, as \citet[5]{linnemann_hints_2018} notice, ``[i]n the current GR [general relativity] era `gravity' is often used synonymously with `spacetime'." \citet{van_raamsdonk_lectures_2017} formulates himself more precisely when he, based on the same results, proposes that ``gravitational dynamics can be seen to emerge directly from the physics of entanglement." This emergence claim seems to fare better than the claim that the metric emerges from entanglement (if we still disregard the asymmetry needed for emergence). Where Section \ref{DeterminationNovelty} questioned whether novelty was satisfied for metric emergence because both the emergent and emergence basis (once fully furnished) include a metric, gravitational dynamics is not included in the emergence basis implied by the result recounted above. Gravitational dynamics or the coupling of the metric to matter through a dynamical equation is precisely what is absent on the CFT side which instead encodes this dynamics in the dynamics of the entanglement structure. Section \ref{Metric} speculated that this difference could also be counted towards satisfying the novelty condition for metric emergence. After all, it is an important difference between the bulk and boundary metric. However, because we have the option to state this difference in terms of `the emergence of gravitational dynamics,' it seems unnecessarily imprecise to state it in terms of `the emergence of the metric' since the latter statement could be misinterpreted as entailing the complete absence of a metric on the CFT side. Insisting on the distinction between the emergence of gravitational dynamics and the emergence of the metric has, at least, the communicative purpose of distinguishing two emergence claims relating to the metric where that about gravitational dynamics more clearly satisfies the novelty condition. 

What is novel, though, is not dynamics in itself. The AdS side and CFT side are descriptions of the same physics. Even though the dynamics is captured by different dynamical equations on each side, the dynamics could even be argued to be, in some sense, the same. What is novel is \textit{gravitational} dynamics, dynamics realized by a particular kind of ontological model that involves the coupling of metric to matter. One might, though, take the debate over the equivalence of general relativity and teleparallel gravity \citep{weatherall_are_2025} and that over the difference between modified gravity and dark matter models \citep{martens_dark_2020} as indications that distinguishing kinds of dynamics by the ontological model that realize them might be questioned. Thus, it is with this caveat that gravitational dynamics on the AdS side is here judged to be novel compared to the CFT side. 

The determination condition, i.e., that the emergence basis determines the emergent, is not satisfied in this case, either, even though \citet{van_raamsdonk_lectures_2017}, for instance, indicates that the emergence basis only contains ``the physics of entanglement." In the result that demonstrates the relation between gravitational dynamics and entanglement dynamics, more elements on the CFT side besides entanglement are still needed to determine the how the bulk metric couples to matter, one of these additional elements being a CFT energy. Thus, van Raamsdonk's claim that gravitational dynamics emerges from entanglement only satisfies the determination condition if it is interpreted as a contrasting emergence claim where some of the emergence basis is left implicit. 

\section{Connectivity from Entanglement}\label{Topology}
\begin{figure}[h]
\begin{center}
\begin{tikzpicture}
\node (I)    at (-3,-0.5)   {I};
\node (II)   at (-7,-0.5)   {II};
\node (III)  at (-5, 2) {IV};
\node (IV)   at (-5,-2) {III};

     \path  (-2,3)  coordinate  (IItop);
     \path  (-2,-3) coordinate (IIbot);
     \path  (-5,0)   coordinate (mid);
     
     \node at (-8,3.3) {$i^+$};
     \node at (-2,3.3) {$i^+$};
     \node at (-8,-3.3) {$i^-$};
     \node at (-2,-3.3) {$i^-$};

\draw (IItop) -- (mid) node[scale=0.7,midway, above, sloped, inner sep=2mm] {$r=r_h$};
\draw (mid) -- (IIbot);
\draw (IItop) -- (IIbot) node[midway, right, inner sep=2mm] {$A$};

 \path      (-8,3)  coordinate (Itop);
 \path      (-8,-3) coordinate (Ibot);
       
\draw  (Itop) -- (Ibot) node[midway, left, inner sep=2mm] {$B$};
\draw (Ibot) -- (mid);
\draw (mid) -- (Itop) node[scale=0.7,midway, above, sloped, inner sep=2mm] {$r=r_h$};

\draw[decorate,decoration=zigzag] (IItop) -- (Itop) node[scale=0.7,midway, above, inner sep=2mm] {$r=0$};

\draw[decorate,decoration=zigzag] (IIbot) -- (Ibot) node[scale=0.7,midway, above, inner sep=2mm] {$r=0$};

\end{tikzpicture}
\caption{\label{penrosethermofield} Penrose diagram of the eternal black hole. The regions I and II lie outside ($r>r_h$) the black hole (region IV). Region III is a white hole. At each point, there is an implicit $d-1$ dimensional sphere that scales as $r^{d-1}$. Adapted from \citet[4]{jaksland_probing_2018}.}
\end{center}
\end{figure}
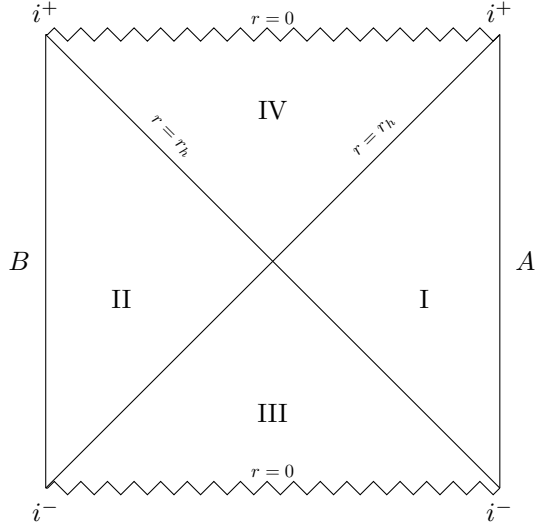

When \citet[2328]{van_raamsdonk_building_2010} in another paper concludes ``that the intrinsically quantum phenomenon of entanglement appears to be crucial for the emergence of classical spacetime geometry," the findings, however, neither concern gravitational dynamics nor the determination of the bulk metric. In the investigation leading up to this conclusion, van Raamsdonk considers a CFT defined on a particular manifold, $A=\mathbb{R} \times S^{(d-1)}$, with Hilbert space $\mathcal{H}_A$ and a copy of this system, i.e., another CFT defined on the distinct but identical manifold, $B$, with the identical Hilbert space $\mathcal{H}_B$. The composite system, in other words, is defined on the manifold $X = A \sqcup B$ (where $\sqcup$ is the disjoint union), and states of the composite system, $\ket{\Psi}$, are in $\mathcal{H}_A \otimes \mathcal{H}_B$. One of the states of the composite system is the product state of two identical energy eigenstates $\ket{\Psi_P}=\ket{E_1^A} \otimes \ket{E_1^B}$. The AdS dual of $\ket{\Psi_P}$, $M_\Psi$, consists of two copies of the AdS dual of $\ket{E_1^A}=\ket{E_1^B}$: ``the product state is dual to the disconnected pair of spacetimes" \citep[2324]{van_raamsdonk_building_2010}. Things change, however, if the composite CFT state is instead a thermofield double state, 
\begin{equation}\label{double}
\ket{\Psi_T}=\frac{1}{\sqrt{Z(\beta)}}\sum_i e^{-\beta E_i / 2} \ket{E_i^A} \otimes \ket{E_i^B}.
\end{equation}
This is a non-separable state, i.e., the two subsystems are entangled. While these subsystems are still defined on distinct manifolds, the thermofield double state, \citet[2324]{van_raamsdonk_building_2010} observes, ``corresponds to the (connected) eternal AdS black hole" (see figure \ref{penrosethermofield} for its Penrose diagram). Based on this, \citet[2325]{van_raamsdonk_building_2010} argues that, ``In this example, classical connectivity arises by entangling the degrees of freedom in the two [CFT] components." Thus, when van Raamsdonk, as quoted above, summarizes this and related results as involving the ``emergence of classical spacetime geometry," `spacetime' here seems to mean ``classical connectivity." 

What van Raamsdonk in turn means by ``classical connectivity" is also somewhat ambiguous, and he never states this more precisely. From the above discussion, the most immediate understanding of ``connectivity" is the topological property of connectedness. This is a property of topological spaces, such as manifolds, that, loosely, captures whether the space is one whole or consists of multiple distinct topological spaces. Formally, a topological space is disconnected if it is the disjoint union of two non-empty, disjoint open subsets. For manifolds, but not topological spaces in general, connectedness is equivalent to path-connectedness. A topological space, $Y$, is path-connected if, for every two points $x,y \in Y$, there is a continuous path between the two points through that space, i.e. if there is a continuous function, $f$, from the unit interval to $Y$ such that $f(0)=x$ and $f(1)=y$. Arguably, path-connectedness aligns better with the notion that van Raamsdonk has in mind. 

Evidently, on the CFT side, there is no such path from a point in $A$ to a point in $B$. Therefore, $X$---the manifold on which the CFT is defined and the asymptotic boundary of the AdS dual---is path-disconnected and, by implication, topologically disconnected. The same holds for the AdS dual of the non-entangled, product state $\ket{\Psi_P}$. The AdS eternal black hole, which is the dual of the entangled, thermofield double state, $\ket{\Psi_T}$, is, by contrast, path-connected. There is a continuous path between every two points of its manifold. Thus, van Raamsdonk's claim seems, more precisely, to be that, for a CFT on $X$, bulk topological (path-)connectivity emerges from entanglement. He does indicate, though, that the scope of the conclusion is not limited to a CFT on $X$ when he summarizes the results as showing that ``the emergence of classically connected spacetimes is intimately related to the quantum entanglement of degrees of freedom in a non-perturbative description of quantum gravity" \citep[2323]{van_raamsdonk_building_2010}, where the ``non-perturbative description" is the CFT description.\footnote{The argument uses the HRT formula to monitor what happens to the area of the extremal codimension-2 surface that separates the two asymptotic boundaries in the eternal black hole when entanglement is removed between the two CFT components. When entanglement is removed, this area goes to zero which \citet[2326]{van_raamsdonk_building_2010} takes as evidence for the AdS manifold becoming disconnected. For more on this argument, see \citet{jaksland_probing_2018} and \citet{bain_rt_2021}.} 

In a companion piece, van Raamsdonk makes more precise what he means by ``intimately related to" when he ``conclude[s] that entanglement is \textit{necessary} for a bulk picture in which the two regions of spacetime are connected" \citep[15, emphasis added]{van_raamsdonk_comments_2010}. The two regions that van Raamsdonk is here referring to are those separated by a boundary-anchored, extremal codimension-2 surface through the bulk. More precisely, the proposal is that there must be entanglement on the CFT side between the regions separated by the endpoints of that extremal surface for the two corresponding bulk regions to be connected. Because "connected" is here a relation between two regions, van Raamsdonk cannot mean the global property of being (path-)connected. Rather, what he means by ``connected" must be that, for every two points, one on each side of the extremal surface, there is a path throug the bulk between those two points. However, if both regions are themselves path-connected, then this does entail that the whole manifold is path-connected. Thus, we can still understand it in terms of the global topological property of path-connectedness when van Raamsdonk proposes that entanglement is necessary for ``the emergence of classically connected spacetimes."\footnote{In the companion piece, van Raamsdonk also states the narrower conclusion for a CFT on $X$ in these slightly different terms: ``If product states correspond to completely disconnected spacetimes, it follows that any spacetime in which the two asymptotic regions are connected must be described by an entangled state" \citep[13]{van_raamsdonk_comments_2010}. By talking about how the ``the two asymptotic regions are connected," van Raamsdonk must here mean that there exists a path in the topological sense between the two regions through the spacetime. The existence of such a path is a question that can be asked with respect to two given subsets of a topological space whereas topological connectivity is a global property of the topological space.}

Assessing this emergence claim promises to be especially clean because the emergence claim only concerns the presence and absence of entanglement rather than the details of the entanglement structure as was the case for the emergence of metric and gravitational dynamics. Again, the question is whether the emergence claim satisfies the determination and novelty conditions. In the general case, van Raamsdonk already states that entanglement is only necessary for the emergence of bulk connectivity, i.e., more elements are needed to determine that the bulk is (path-)connected. Indeed, when van Raamsdonk writes that ``entanglement appears to be crucial for the emergence of classical spacetime geometry," he might be interpreted as clearly signaling that this is a contrasting emergence claim. Entanglement is not the only element in the emergence basis, but it is ``crucial," by which he might mean that entanglement is the element of the emergence basis that is worth singling out.\footnote{\citet[section 3]{bain_rt_2021} elaborates on this point by showing that boundary entanglement underdetermines the bulk topology.}

For a CFT on $X$, however, van Raamsdonk says that connectivity ``arises" from entanglement between the degrees of freedom in $A$ and $B$ which might suggest that entanglement, in this case, is the only element in the emergence basis. However, also in this case, the determination condition is not satisfied. Because the parameter $\beta$ in the thermofield double state is inversely proportional to the temperature of the black hole and an AdS black hole has a finite smallest temperature, the black hole will cease to exist for large but finite $\beta$. For large but finite $\beta$, the AdS dual is instead two copies of thermal AdS \citep[848]{chen_holographic_2019}, i.e., a (path-)disconnected spacetime. However, for all finite $\beta$, $\ket{\Psi_T}$ remains an entangled state. Thus, the degrees of freedom in $A$ and $B$ can be entangled while the dual spacetime is disconnected. Even for a CFT defined on $X=A \sqcup B$, entanglement can at most be a necessary condition for connectivity. Also here, more elements than entanglement must be included in the emergence basis to satisfy the determination condition.

What extra elements to include in the emergence basis remains, however, unanswered for both a CFT on $X$ and in the general case. Something can still be said about the novelty condition by considering the full CFT side which must be sufficient to determine connectivity, though not all of it might be necessary. In the case of a CFT defined on the (path-)disconnected manifold, $X$, in the state $\ket{\Psi_T}$, there is a rather clear difference in properties between the emergent AdS side and the CFT side which contains the emergence basis. The AdS side is (path-)connected whereas the CFT side is not. Thus, (path-)connectivity is a new property of the emergent topological space that is not shared by the topological space of the emergence basis. In one sense, though, there is already much connectivity on the CFT side. After all, both $A$ and $B$ are (path-)connected topological spaces. The global property of being path-disconnected is a logical compound property consisting of a conjunction of instances of the relation of there being a path between two non-empty subsets of a topological space. On the CFT side, there are numerous instances of such paths. There just happens not to be any such paths between subsets of $A$ and subsets of $B$. In other words, only the global logical compound property is new compared to the emergence basis whereas instances of the component relation of this logical compound property are already present in the emergence basis. 

One might argue that the requirement of qualitative difference in the novelty condition precisely seeks to rule out that trivially complex properties like logical compound properties can be considered emergent. Under this assumption, the novelty condition is not satisfied for emergent path-connectedness if the existence of a path between two subsets is part of the CFT emergence basis.\footnote{One cannot avoid this problem by refocusing the emergence claim on path-\textit{dis}connectedness because entanglement is not necessary for path-disconnectedness. Disconnectedness does not supervene on entanglement in violation of the determination condition.} And this will be the case unless the AdS side can be path-connected even when no paths exist between any subsets on the CFT side, i.e., in a case where the CFT side is totally path-disconnected. To my knowledge, the literature contains no such result, though, and investigating it further is beyond the scope here. What can be said, therefore, is that more than entanglement is needed on the CFT side to determine topological path-connectivity on the AdS side, and it remains unclear whether novelty is satisfied by the fully furnished emergence basis.

\section{Conclusion}
The AdS/CFT literature contains numerous variants of the claim that spacetime emerges from entanglement. This paper has shown that behind these generalizing statements hide three distinct claims about the emergence of different spatiotemporal aspects from entanglement: metric, gravitational dynamics, and topological connectivity. This disambiguation is not meant to suggest that this is a cause for any confusion where physicists, for instance, mistakenly rely on results pertaining to the bulk metric to make derivations about gravitational dynamics. Rather, that there are three distinct claims being made with the same words is of consequence to those physicists and philosophers alike who draw the big picture implications of these individual results. Despite being advertised with the same terminology, these results are not about the same spatiotemporal aspects and they can therefore not just be grouped as evidence for one unanimous conclusion, though they, of course, point in similar directions. 

Furthermore, this paper has shown that more than entanglement is necessary in the emergence basis for all of the three distinct emergence claims. In the terminology adopted here, all of them should be understood as contrasting emergence claims where some of the emergence basis is left implicit. For the claim that gravitational dynamics emerges from entanglement, this is of little consequence. For the claim that the metric emerges from entanglement, however, explicating the implicit elements of the emergence basis reveals that it already contains a metric. Thus, this claim might not satisfy the condition for emergence that the emergent should be novel as compared to the emergence basis. If, by spacetime, we mean metric (plus manifold), then spacetime is present on both the AdS and CFT side. For the claim that connectivity emerges from entanglement, it remains unknown how to complete the emergence basis, but if this emergence basis includes paths between subsets, it was argued that this emergence claim might not satisfy novelty either. 

This paper can, in other words, be seen as giving three warnings to those who are interested in the philosophical or big picture implications of the spacetime from entanglement hypothesis: (a) `Spacetime' is used to refer to different spatiotemporal aspects in different contexts; (b) The emergence claims should, at least in all the cases studied here, be understood as contrasting emergence claims which leaves parts of the emergence basis implicit; (c) The implicit emergence basis might include elements that are not sufficiently qualitatively different from the emergent to actually qualify as possible instances of emergence.

\section*{Funding}
The research is funded by the Carlsberg Foundation, grant number CF22-1428, and the John Templeton Foundation, grant ID 62851. The opinions expressed in this publication are those of the author and do not necessarily reflect the views of the John Templeton Foundation.

\printendnotes

\sloppy

\section*{References}
Akers, Chris, Venkatesa Chandrasekaran, Stefan Leichenauer, Adam Levine, and Arvin Shahbazi Moghaddam. 2020. ``Quantum null energy condition, entanglement wedge nesting, and quantum
focusing." \textit{Physical Review D} 101 (2): 025011. \href{https://doi.org/10.1103/PhysRevD.101.025011}{https://doi.org/10.1103/PhysRevD.101.025011}.

Bain, Jonathan. 2021. ``The RT formula and its discontents: spacetime and entanglement." \textit{Synthese} 198 (12): 11833–11860. \href{https://doi.org/10.1007/s11229-020-02836-4}{https://doi.org/10.1007/s11229-020-02836-4}.

Bantilan, Hans, Pau Figueras, and Lorenzo Rossi. 2021. ``Cauchy evolution of asymptotically global AdS spacetimes with no symmetries." \textit{Physical Review D} 103 (8): 086006. \href{https://doi.org/10.1103/PhysRevD.103.086006}{https://doi.org/10.1103/PhysRevD.103.086006}.

Bao, Ning, ChunJun Cao, Sebastian Fischetti, and Cynthia Keeler. 2019. ``Towards bulk metric reconstruction from extremal area variations." \textit{Classical and Quantum Gravity} 36 (18): 185002. \href{https://doi.org/10.1088/1361-6382/ab377f}{https://doi.org/10.1088/1361-6382/ab377f}.

Bao, Ning, ChunJun Cao, Sebastian Fischetti, Jason Pollack, and Yibo Zhong. 2021. ``More of the bulk from extremal area variations." \textit{Classical and Quantum Gravity} 38 (4): 047001. \href{https://doi.org/10.1088/1361-6382/abcfd0}{https://doi.org/10.1088/1361-6382/abcfd0}.

Baron, Sam. 2020. ``The curious case of spacetime emergence." \textit{Philosophical Studies} 177 (8): 2207–2226. \href{https://doi.org/10.1007/s11098-019-01306-z}{https://doi.org/10.1007/s11098-019-01306-z}.

Bianchi, Eugenio, and Robert C. Myers. 2014. ``On the architecture of spacetime geometry." \textit{Classical and Quantum Gravity} 31 (21): 214002.

Bishop, Robert C. 2022. ``Contextual Emergence: Constituents, Context and Meaning." In \textit{From Electrons to Elephants and Elections: Exploring the Role of Content and Context}, edited by Shyam Wuppuluri and Ian Stewart, 243–256. Cham: Springer International Publishing. \href{https://doi.org/10.1007/978-3-030-92192-7 15}{https://doi.org/10.1007/978-3-030-92192-7 15}.

Bishop, Robert C., and Harald Atmanspacher. 2006. ``Contextual Emergence in the Description of Properties." \textit{Foundations of Physics} 36 (12): 1753–1777. \href{https://doi.org/10.1007/s10701-006-9082-8}{https://doi.org/10.1007/s10701-006-9082-8}.

Burda, Philipp, Ruth Gregory, and Akash Jain. 2019. ``Holographic reconstruction of bubble spacetimes." \textit{Physical Review} D 99 (2): 026003. \href{https://doi.org/10.1103/PhysRevD.99.026003}{https://doi.org/10.1103/PhysRevD.99.026003}.

Butterfield, Jeremy. 2011a. ``Emergence, Reduction and Supervenience: A Varied Landscape." \textit{Foundations of Physics} 41 (6): 920–959.

-----. 2011b. ``Less is Different: Emergence and Reduction Reconciled." \textit{Foundations of Physics} 41 (6): 1065–1135.

Casini, Horacio, and Marina Huerta. 2022. ``Lectures on entanglement in quantum field theory." \textit{PoS TASI2021}:002.

Chen, Bin. 2019. ``Holographic Entanglement Entropy: A Topical Review." \textit{Communications in Theoretical Physics} 71 (7): 837. \href{https://doi.org/10.1088/0253-6102/71/7/837}{https://doi.org/10.1088/0253-6102/71/7/837}.

Compère, Geoffrey, and Donald Marolf. 2008. ``Setting the boundary free in AdS/CFT." \textit{Classical and Quantum Gravity} 25 (19): 195014. \href{https://doi.org/10.1088/0264-9381/25/19/195014}{https://doi.org/10.1088/0264-9381/25/19/195014}.

Crowther, Karen. 2015. ``Decoupling emergence and reduction in physics." \textit{European Journal for Philosophy of Science} 5 (3): 419–445. \href{https://doi.org/10.1007/s13194-015-0119-8}{https://doi.org/10.1007/s13194-015-0119-8}.

-----. 2021. ``As below, so before: ‘synchronic’ and ‘diachronic’ conceptions of spacetime emergence." \textit{Synthese} 198 (8): 7279–7307. \href{https://doi.org/10.1007/s11229-019-02521-1}{https://doi.org/10.1007/s11229-019-02521-1}.

-----. 2022. ``Spacetime Emergence: Collapsing the Distinction Between Content and Context?" In \textit{From Electrons to Elephants and Elections: Exploring the Role of Content and Context}, edited by Shyam Wuppuluri and Ian Stewart, 379–402. Cham: Springer International Publishing.

De Haro, Sebastian, and Jeremy Butterfield. 2018. ``A Schema for Duality, Illustrated by Bosonization." In \textit{Foundations of Mathematics and Physics One Century After Hilbert: New Perspectives}, edited by Joseph Kouneiher, 305–376. Cham: Springer International Publishing. \href{https://doi.org/10.1007/978-3-319-64813-2 12}{https://doi.org/10.1007/978-3-319-64813-2 12}.

Dieks, Dennis, Jeroen van Dongen, and Sebastian de Haro. 2015. ``Emergence in holographic scenarios for gravity." \textit{Studies in History and Philosophy of Science Part B: Studies in History and Philosophy of Modern Physics} 52: 203–216.

Engelhardt, Netta, and Sebastian Fischetti. 2019. ``Surface theory: the classical, the quantum, and the holographic." \textit{Classical and Quantum Gravity} 36 (20): 205002. \href{https://doi.org/10.1088/1361-6382/ab3bda}{https://doi.org/10.1088/1361-6382/ab3bda}.

Faulkner, Thomas, Monica Guica, Thomas Hartman, Robert C. Myers, and Mark Van Raamsdonk. 2014. ``Gravitation from entanglement in holographic CFTs." \textit{Journal of High Energy Physics} 2014 (3): 51.

Faulkner, Thomas, Felix M. Haehl, Eliot Hijano, Onkar Parrikar, Charles Rabideau, and Mark Van Raamsdonk. 2017. ``Nonlinear gravity from entanglement in conformal field theories." \textit{Journal of High Energy Physics} 2017 (8): 57.

Haro, Sebastian de. 2017. ``Dualities and emergent gravity: Gauge/gravity duality." \textit{Studies in History
and Philosophy of Science Part B: Studies in History and Philosophy of Modern Physics} 59:109–125.

Hubeny, Veronika E. 2015. ``The AdS/CFT correspondence." \textit{Classical and Quantum Gravity} 32 (12): 124010.

Hubeny, Veronika E, Mukund Rangamani, and Simon F. Ross. 2005. ``Causally Pathological Space–Times are Physically Relevant." \textit{International Journal of Modern Physics} D 14 (12): 2227–2231. \href{https://doi.org/10.1142/S0218271805007760}{https://doi.org/10.1142/S0218271805007760}.

Hubeny, Veronika E, Tadashi Takayanagi, and Mukund Rangamani. 2007. ``A covariant holographic entanglement entropy proposal." \textit{Journal of High Energy Physics} 0707 (062).

Huggett, Nick, and Christian Wüthrich. 2025. \textit{Out of Nowhere: The Emergence of Spacetime in
Quantum Theories of Gravity}. Oxford: Oxford University Press.

Jaksland, Rasmus. 2018. ``Probing spacetime with a holographic relation between spacetime and entanglement." \href{https://philsci-archive.pitt.edu/15415/}{https://philsci-archive.pitt.edu/15415/}.

-----. 2021. ``Entanglement as the world-making relation: Distance from entanglement." \textit{Synthese} 198:9661–9693. \href{https://doi.org/10.1007/s11229-020-02671-7}{https://doi.org/10.1007/s11229-020-02671-7}.

Jaksland, Rasmus, and Niels Linnemann. 2020. ``Holography without holography: How to turn inter-representational into intra-theoretical relations in AdS/CFT." \textit{Studies in History and Philosophy of Science Part B: Studies in History and Philosophy of Modern Physics} 71:101–117. \href{https://doi.org/10.1016/j.shpsb.2020.04.007}{https://doi.org/10.1016/j.shpsb.2020.04.007}.

-----. 2024. ``On the Non-identity Causal Theory of Spacetime from Causal Set Theory." \textit{Erkenntnis}, \href{https://doi.org/10.1007/s10670-024-00836-1}{https://doi.org/10.1007/s10670-024-00836-1}.

Jaksland, Rasmus, and Kian Salimkhani. 2023. ``The many problems of spacetime emergence in quantum gravity." \textit{The British Journal for the Philosophy of Science}, \href{https://doi.org/10.1086/727052}{https://doi.org/10.1086/727052}.

Kamal, Helia, and Geoffrey Penington. 2019. ``The Ryu-Takayanagi Formula from Quantum Error Correction: An Algebraic Treatment of the Boundary CFT." \href{https://arxiv.org/abs/1912.02240.24}{https://arxiv.org/abs/1912.02240.24}

Lashkari, Nima, Michael B. McDermott, and Mark Van Raamsdonk. 2014. ``Gravitational dynamics from entanglement ``thermodynamics"." \textit{Journal of High Energy Physics} 2014 (4): 195. \href{https://doi.org/https://doi.org/10.1007/JHEP04(2014)195}{https://doi.org/https://doi.org/10.1007/JHEP04(2014)195}.

Le Bihan, Baptiste. 2018. ``Priority Monism beyond Spacetime." \textit{Metaphysica} 19 (1): 95–111. \href{https://doi.org/10.1515/mp-2018-0005}{https://doi.org/10.1515/mp-2018-0005}.

Linnemann, Niels, and Manus R. Visser. 2018. ``Hints towards the emergent nature of gravity." \textit{Studies in History and Philosophy of Science Part B: Studies in History and Philosophy of Modern Physics} 64:1–13. \href{https://doi.org/10.1016/j.shpsb.2018.04.001}{https://doi.org/10.1016/j.shpsb.2018.04.001}.

Luu, Thomas, and Ulf-G. Meißner. 2021. ``On the Topic of Emergence from an Effective Field Theory Perspective." In \textit{Top-Down Causation and Emergence}, edited by Jan Voosholz and Markus Gabriel, 101–114. Cham: Springer International Publishing. \href{https://doi.org/10.1007/978-3-030-71899-2 5}{https://doi.org/10.1007/978-3-030-71899-2 5}.

Marolf, Donald, William Kelly, and Sebastian Fischetti. 2014. ``Conserved Charges in Asymptotically (Locally) AdS Spacetimes." In \textit{Springer Handbook of Spacetime}, edited by Abhay Ashtekar and Vesselin Petkov, 381–407. Berlin, Heidelberg: Springer Berlin Heidelberg.

Martens, Niels C.M., and Dennis Lehmkuhl. 2020. ``Dark matter = modified gravity? Scrutinising the spacetime–matter distinction through the modified gravity/ dark matter lens." \textit{Studies
in History and Philosophy of Science Part B: Studies in History and Philosophy of Modern Physics} 72:237–250. \href{https://doi.org/10.1016/j.shpsb.2020.08.003}{https://doi.org/10.1016/j.shpsb.2020.08.003}.

Matarese, Vera. 2019. ``Loop Quantum Gravity: A New Threat to Humeanism? Part I: The Problem of Spacetime." \textit{Foundations of Physics} 49 (3): 232–259. \href{https://doi.org/10.1007/s10701-019-00242-6}{https://doi.org/10.1007/s10701-019-00242-6}.

Matsubara, Keizo, and Lars-Göran Johansson. 2018. ``Spacetime in String Theory: A Conceptual Clarification." \textit{Journal for General Philosophy of Science} 49 (3): 333–353. \href{https://doi.org/10.1007/s10838-018-9423-2}{https://doi.org/10.1007/s10838-018-9423-2}.

Minces, Pablo, and Victor O. Rivelles. 2000. ``Scalar field theory in the AdS/CFT correspondence revisited." \textit{Nuclear Physics B} 572 (3): 651–669. \href{https://doi.org/10.1016/S0550-3213(99)00833-0}{https://doi.org/10.1016/S0550-3213(99)00833-0}.

Ney, Alyssa. 2021. ``From Quantum Entanglement to Spatiotemporal Distance." In \textit{Philosophy Beyond Spacetime}, edited by Christian Wüthrich, Baptiste Le Bihan, and Nick Huggett, 78–102.
New York: Oxford University Press.

Nomura, Yasunori, Nico Salzetta, Fabio Sanches, and Sean J. Weinberg. 2016. ``Spacetime equals entanglement." \textit{Physics Letters B} 763:370–374. \href{https://doi.org/10.1016/j.physletb.2016.10.045}{https://doi.org/10.1016/j.physletb.2016.10.045}.

Palacios, Patricia. 2022. \textit{Emergence and Reduction in Physics}. Elements in the Philosophy of Physics. Cambridge: Cambridge University Press. \href{https://doi.org/10.1017/9781108901017}{https://doi.org/10.1017/9781108901017}.

Rangamani, Mukund, and Tadashi Takayanagi. 2017. \textit{Holographic entanglement entropy}. Cham, Switzerland: Springer.

Rickles, Dean. 2013. ``AdS/CFT duality and the emergence of spacetime." \textit{Studies in History and Philosophy of Science Part B: Studies in History and Philosophy of Modern Physics} 44 (3):
312–320.

Ryu, Shinsei, and Tadashi Takayanagi. 2006. ``Holographic Derivation of Entanglement Entropy from the anti-de Sitter Space/Conformal Field Theory Correspondence." \textit{Physical Review Letter} 96 (18): 181602.

Swingle, Brian. 2018. ``Spacetime from Entanglement." \textit{Annual Review of Condensed Matter Physics} 9 (1): 345–358. \href{https://doi.org/10.1146/annurev-conmatphys-033117-054219}{https://doi.org/10.1146/annurev-conmatphys-033117-054219}.

Swingle, Brian, and Mark Van Raamsdonk. 2014. ``Universality of Gravity from Entanglement." \href{https://arxiv.org/abs/arXiv:1405.2933v1}{https://arxiv.org/abs/arXiv:1405.2933v1}.

Taylor, Marika, and William Woodhead. 2016. ``Renormalized entanglement entropy." \textit{Journal of High Energy Physics} 2016 (8): 165. \href{https://doi.org/10.1007/JHEP08(2016)165}{https://doi.org/10.1007/JHEP08(2016)165}.

Teh, Nicholas J. 2013. ``Holography and emergence." \textit{Studies in History and Philosophy of Science Part B: Studies in History and Philosophy of Modern Physics} 44 (3): 300–311.

Truesdell, C. 1975. ``Early Kinetic Theories of Gases." \textit{Archive for History of Exact Sciences} 15 (1): 1–66.

Van Raamsdonk, Mark. 2010a. ``Building up spacetime with quantum entanglement." \textit{General Relativity and Gravitation} 42 (10): 2323–2329.

------. 2010b. ``Comments on quantum gravity and entanglement." \href{https://arxiv.org/abs/0907.2939v2}{https://arxiv.org/abs/0907.2939v2}.

------. 2017. ``Lectures on Gravity and Entanglement." In \textit{New Frontiers in Fields and Strings}, 297–351. WORLD SCIENTIFIC. \href{https://doi.org/10.1142/9789813149441_0005}{https://doi.org/10.1142/9789813149441\_0005}.

------. 2020. ``Spacetime from bits." \textit{Science} 370 (6513): 198–202. \href{https://doi.org/10.1126/science.aay9560}{https://doi.org/10.1126/science.aay9560}.

Wall, Aron C. 2014. ``Maximin surfaces, and the strong subadditivity of the covariant holographic entanglement entropy." \textit{Classical and Quantum Gravity} 31 (22): 225007. \href{https://doi.org/10.1088/0264-9381/31/22/225007}{https://doi.org/10.1088/0264-9381/31/22/225007}.

Wayne, Andrew, and Michal Arciszewski. 2009. ``Emergence in Physics." \textit{Philosophy Compass} 4 (5):
846–858. \href{https://doi.org/10.1111/j.1747-9991.2009.00239.x}{https://doi.org/10.1111/j.1747-9991.2009.00239.x}.

Weatherall, James Owen, and Helen Meskhidze. 2025. ``Are General Relativity and Teleparallel Gravity Theoretically Equivalent?" \textit{Philosophy of Physics,} \href{https://doi.org/10.31389/pop.152}{https://doi.org/10.31389/pop.152}.

Wüthrich, Christian. 2020. ``When the actual world is not even possible." In \textit{The foundation of reality: Fundamentality, space, and time}, edited by David Glick, George Darby, and Anna Marmodoro. Oxford: Oxford University Press.

Yngvason, Jakob. 2005. ``The role of type III factors in quantum field theory." \textit{Reports on Mathematical Physics} 55 (1): 135–147. \href{https://doi.org/10.1016/S0034-4877(05)80009-6}{https://doi.org/10.1016/S0034-4877(05)80009-6}.


\end{document}